\documentclass[journal,twocolumn]{IEEEtran}
\usepackage{epsfig,makeidx,color}
\usepackage{amsmath,amssymb,bbm}
\usepackage{cite,graphicx,lipsum}
\usepackage{enumerate}
\usepackage[switch,pagewise]{lineno}
\usepackage{hyperref}
\hypersetup{
        colorlinks = true,
        citecolor=blue,
}
\def\cH{{\cal H}}
\def\cK{{\cal K}}

\def\cQ{{\cal Q}}

\def\rH{{\rm H}}
\def\rT{{\rm T}}

\def\uP{{\mathbb P}}

\def\uC{{\mathbb C}}
\def\uE{{\mathbb E}}

\DeclareMathOperator*{\argmin}{\arg\!\min}
\DeclareMathOperator*{\argmax}{\arg\!\max}

\def\deft{ \buildrel \triangle \over = }

\def\be{ \begin{equation} }
\def\ee{ \end{equation} }
\def\bea{ \begin{eqnarray} }
\def\eea{ \end{eqnarray} }

\def\bx{{\bf x}}
\def\by{{\bf y}}
\def\bc{{\bf c}}

\def\bb{{\bf b}}

\def\bg{{\bf g}}

\def\bs{{\bf s}}
\def\ba{{\bf a}}
\def\br{{\bf r}}
\def\bu{{\bf u}}

\def\bn{{\bf n}}
\def\bh{{\bf h}}

\def\bz{{\bf z}}

\def\bv{{\bf v}}

\def\bA{{\bf A}}

\def\bC{{\bf C}}

\def\bI{{\bf I}}

\def\bN{{\bf N}}

\def\bP{{\bf P}}

\def\bR{{\bf R}}
\def\bW{{\bf W}}
\def\bU{{\bf U}}

\def\bY{{\bf Y}}
\def\bZ{{\bf Z}}

\def\b0{{\bf 0}}
\def\bPhi{{\bf \Phi}}

\def\cC{{\cal C}}
\def\cL{{\cal L}}
\def\cJ{{\cal J}}

\def\cB{{\cal B}}

\def\cQ{{\cal Q}}

\def\cN{{\cal N}}

\def\sD{{\sf D}}

\ifCLASSOPTIONonecolumn
  \newcommand{\figwidth}{0.50\columnwidth}
  
\else
  \newcommand{\figwidth}{0.90\columnwidth}
  
\fi

\begin{document}

%\title{On Layered Preambles based on NOMA for 
%Two Different Types of Devices in MTC}
\title{Random Access with Layered Preambles based on NOMA 
for Two Different Types of Devices in MTC}

\author{Jinho Choi\\
\thanks{The author is with
the School of Information Technology,
Deakin University, Geelong, VIC 3220, Australia
(e-mail: jinho.choi@deakin.edu.au).}}

\maketitle
\begin{abstract}
In machine-type communication (MTC), random 
access has been employed for 
a number of devices and sensors to access
uplink channels using a pool of preambles. To support
different priorities due to 
various quality-of-service (QoS) requirements,
random access can be generalized with multiple pools, 
which may result in low spectral efficiency.
In this paper, for high spectral efficiency,
random access with layered preambles
(RALP) is proposed to support
devices with two different priorities based
on the notion of power-domain 
non-orthogonal multiple access (NOMA).
In RALP, two groups of devices,
namely type-1 and type-2 devices,
are supported with different priorities,
where type-1 devices have higher priority than type-2 devices.
Closed-form expressions are derived for the detection performance
of preambles transmitted by type-1 devices, which can be used for
a certain performance guarantee of type-1 devices of high priority.
Low-complexity preamble detection methods are also discussed.
\end{abstract}


\ifCLASSOPTIONonecolumn
\baselineskip 28pt
\fi

\section{Introduction}

The Internet of Things (IoT) is a network
of things that are connected for 
a number of applications including 
smart cities and factories \cite{Gubbi13} \cite{Kim16}.
To support the connectivity,
a number of different approaches have
been proposed \cite{Ding_20Access}.
For example, 
in \cite{Qadir18}, low-power wide area networks (LPWAN)
are studied to support devices with long range communications
in unlicensed bands. 
Cellular IoT using
machine-type communication (MTC)
\cite{3GPP_MTC} \cite{3GPP_NBIoT}
is also considered to support the connectivity of IoT
devices and sensors in cellular systems
\cite{Shar15}.
In \cite{Mang16},
a deployment study of narrowband IoT (NB-IoT)
\cite{3GPP_NBIoT} is presented 
for IoT applications 
with sensors and devices deployed 
over a large area within a cellular system.

Due to sparse activity and sporadic traffic of devices and sensors
in MTC \cite{Laner15}, random access is used
to keep signaling overhead low,
and various random access
schemes with a set of preambles 
are studied in handshaking process to establish connections
\cite{3GPP_MTC} \cite{3GPP_NBIoT}
\cite{Chang15} \cite{Choi16}.

To support a large number of devices with limited bandwidth,
non-orthogonal preambles can be used and the detection
of non-orthogonal preambles
can be carried out using multiuser detection approaches
\cite{VerduBook} \cite{ChoiJBook2}.
Since a fraction of devices are active at a time,
the sparse user activity can be taken into account
to design multiuser detectors \cite{Zhu11} \cite{Applebaum12}
\cite{Schepker15}.
The sparse user activity can be represented by
a sparse vector so that the resulting random 
access can be seen as a sparse signal recovery problem
in the context of compressive sensing (CS) 
\cite{Eldar12}, which is called 
\emph{compressive random access} \cite{Wunder14}
\cite{Wunder15}.
Compressive random access can be used
for grant-free random access \cite{Choi17IoT}
\cite{Choi19a} and be combined with
massive multiple-input multiple-output (MIMO)
\cite{Carvalho17} \cite{Bjornson17}
\cite{Liu18} \cite{Ding20b}.

Most compressive random access schemes use a pool of preambles
and any device that has data to transmit is to randomly choose
one preamble from the pool and transmit it.
Although preamble collisions (when
multiple devices choose the same preamble) happen,
this is a way to support a large number of devices
with a limited number of preambles (due to limited bandwidth),
while it is also possible to assign unique sequences
to devices \cite{Choi_18Feb}.
Since there is only one pool, each device
may have almost equal chance to be connected. For example,
each active device has the same probability of preamble collision.
As a result, no priority is introduced
in conventional compressive random access.
However, as in \cite{Laner15} \cite{Li17} \cite{Li17_S} \cite{Zhang19},
devices can have different priorities
depending on their 
Quality-of-Service (QoS) requirements.

To support different priorities, in this paper, we consider 
random access with layered preambles (RALP)
based on the notion of power-domain non-orthogonal
multiple access (NOMA) \cite{Ding_CM} \cite{Choi17_ISWCS}.
In particular, the main contributions are it is assumed that 
there are two different types of devices
in terms of priority, namely type-1 and type-2 devices,
where type-1 devices have higher priority than type-2
devices, while the number of active type-1 devices is much fewer than 
that of active type-2 devices.
In RALP, it is aimed that the probability
of detection errors of type-1 devices
is to be sufficiently low, while that of type-2 devices is arbitrary.
{\color{blue}
In summary, the main contributions of the paper
are as follows: {\it i)} using layered preambles based on 
power-domain NOMA,
a random access scheme to support two different types of devices
is proposed; 
{\it ii)} a low-complexity preamble detection approach is derived using 
successive interference cancellation (SIC) and a well-known
machine learning algorithm, i.e., a variational inference 
(VI) algorithm;
{\it iii)} closed-form expressions for 
preamble detection error probabilities of type-1 devices
are derived.
}

%In particular, we assume that one group of devices,
%which are called type-1 devices,
%requires a low access delay, while the other group
%of devices, which are typical devices
%and called type-2 devices,
%do not have any specific QoS in terms of access delay.

There are a number of related works. For example,
in \cite{Choi19a},
layered preambles are also considered using the notion of
power-domain NOMA, while different priorities are not
taken into account.
In terms of supporting two different priorities in MTC,
\cite{Li17_S} is the most related work,
which mainly focuses on dynamic resource allocation
and user barring without using layered preambles.
In fact, since RALP in this paper 
can provide different priorities with
layered preambles and different detection performance,
it can be used within
dynamic resource allocation and user barring schemes,
which can be seen as a further work.

Note that there are also other NOMA-based random access approaches
where each device has a unique sequence (as a result, they
do not need to use a shared pool of preambles as in MTC).
For example, in \cite{Choi19a} and
\cite{Yuan20}, non-orthogonal Gaussian and low-density signatures,
respectively, are used to improve the spectral efficiency in random 
access and joint channel estimation and detection is considered.
In \cite{Zhang20}, as in \cite{Yuan20}, 
non-orthogonal Gaussian and low-density signatures
are used for unique devices' signature in random access,
where a Bayesian receiver is designed.

The rest of the paper is organized as follows.
In Section~\ref{S:SM}, the system model
is presented. In Section~\ref{S:PN},
RALP is introduced using the notion of
power-domain NOMA to support two different types of devices
with single resource block. 
In Sections~\ref{S:PD1}
and \ref{S:PD2}, the detection methods 
are studied for RALP with performance analysis
for devices of high priority.
Simulation results are presented in Section~\ref{S:Sim}.
Finally, we conclude the paper with remarks
in Section~\ref{S:Con}.

\subsubsection*{Notation}
Matrices and vectors are denoted by upper- and lower-case
boldface letters, respectively.
The superscripts $\rT$ and $\rH$
denote the transpose and complex conjugate, respectively.
The $p$-norm of a vector $\ba$ is denoted by $|| \ba ||_p$
(If $p = 2$, the norm is denoted by $||\ba||$ without
the subscript).
$\uE[\cdot]$
and ${\rm Var}(\cdot)$
denote the statistical expectation and variance, respectively.
$\cC \cN(\ba, \bR)$
represents the distribution of
circularly symmetric complex Gaussian (CSCG)
random vectors with mean vector $\ba$ and
covariance matrix $\bR$.
The Q-function is given by
$\cQ(x) = \int_x^\infty \frac{1}{\sqrt{2 \pi} } e^{- \frac{t^2}{2} } dt$.

\section{System Model}	\label{S:SM}

In this section, the system model
is presented with two different types of 
devices.

\subsection{Random Access for MTC}

Consider a system that consists of a BS and 
a large number of devices that are synchronized for MTC. 
Suppose that a fraction of devices are active
at a time and use random access 
to establish connections to transmit their data 
(e.g., random
access channel (RACH) procedure 
in the long-term evolution advanced (LTE-A) systems \cite{3GPP_MTC}).
For random access, we assume that a common pool of preambles
is used \cite{3GPP_MTC} \cite{3GPP_NBIoT}.
A device that has data packets to transmit, which is called
an active device, randomly chooses 
a preamble from the pool and transmits it to the BS
(through physical random access channel (PRACH) in RACH procedure)
which is the first step of a handshaking process
to establish connection in most MTC schemes (e.g.,
\cite{3GPP_MTC}).
Due to multiple active devices that choose
the same preamble, there exist preamble collisions and
this step can be seen as contention-based access.
There are few more steps 
in the handshaking process
to finally allocate dedicated uplink (data) channels
(which are physical uplink shared channel (PUSCH)
in RACH procedure) 
to active devices so that they can transmit their data packets
to the BS.

Note that it is possible that each device
can have a unique preamble as a signature sequence
so that the BS can identify different devices with their
unique preambles. However, since
the number of devices can be too large to have unique preambles,
devices can share 
a common pool of preambles provided
that devices' activity is sparse (i.e., only a fraction of
devices are active at a time)
at the cost of preamble collisions.

\subsection{Different Types of Devices}

%For devices of equal priority, 
%one pool of preambles is sufficient for random access.
%However, if there are 
%devices of different priorities
%\cite{Li17}, we may need
%to have different pools of preambles.

For the simplicity, we only consider two 
different types of devices in this paper,
which are referred to as
type-1 and type-2 devices,
as follows:
\begin{itemize}
\item Type-1 devices: They require a short access delay,
while the number of them is much fewer than
that of type-2 devices.

\item Type-2 devices: They do not 
require any constraint on access delay.
%while a sufficiently high throughput is expected.
%Here, the throughput is the average number of 
%devices that successfully transmit their preambles.
\end{itemize}
According to \cite{Li17_S},
type-1 and type-2 devices can
be seen as delay-sensitive and delay-tolerant devices,
respectively.

In order to support two different types of devices,
two different (orthogonal) radio resource blocks (RBs)
can be allocated. For each type of devices, 
a pool of preambles can be associated with an RB.
This is the case to build two different access systems.
Note that, with one RB, a pool of preambles
can be dynamically divided into two sub-pools of preambles to support
two types of devices 
with different probabilities of preamble collisions
as in \cite{Li17_S}. 
In this case, in order to have a sufficiently large
number of preambles, a wide system bandwidth might be required,
which may result in a low spectral efficiency.

In fact, the access delay depends on not only
preamble collisions, but also preamble detection errors\footnote{There
can be detection errors due to 
channel fading and/or the noise at the BS.},
since an active device may re-transmit another preamble
if the preamble transmitted previously is not detected
(due to either collision or detection error).
Thus, for a short access delay,
both the probabilities of preamble collisions and detection
errors have to be low.
This implies that it is also necessary to take into account 
the probability of preamble detection errors to support
different priorities, which is not 
considered in \cite{Li17_S}.

In this paper, based on the notion of power-domain NOMA,
we design layered preambles with one RB 
to support different priorities between type-1 and type-2 devices
with a high spectral efficiency
in terms of the probability of preamble detection errors,
which will be discussed in Section~\ref{S:PN}.

%For example, one RB can be assigned
%to transmit preambles of length $L_1$ to type-1 devices. 
%Another RB can be assigned to type-2 devices that 
%are to transmit preambles of length $L_2$.
%If the bandwidth of each RB is proportional to the 
%length of preamble, the total bandwidth becomes
%\be
%W_{\rm tot} \propto  L_1 + L_2.
%	\label{EQ:Wtot}
%\ee
%In other words, the required bandwidth 
%might be proportional to the total number of preambles.

\section{Power-domain NOMA for Layered Preambles with Single RB}
	\label{S:PN}

In this section, we propose layered 
preambles to support two different types of devices
with one RB based on the notion of power-domain NOMA.
%of different priorities (in 

\subsection{Alltop Sequences for Layered Preambles}

For different priorities,
it is necessary to ensure type-1 devices have
a better performance of preamble detection 
than type-2 devices,
which leads to a short access delay.
To this end, while there can be a number of different ways,
we apply power-domain NOMA to certain sequences
of good cross-correlations for preambles.

Suppose that an RB is allocated
to support two different types of devices.
Let $N$ be the length of preamble sequences for a given RB.
Since the system bandwidth is proportional to $N$,
as long as $N$ is fixed, the system bandwidth is fixed.
If all the preambles are orthogonal, the total number of
preambles, denoted by $L$, is equal to $N$. 
However, if non-orthogonal preambles are allowed, $L$ can be 
larger than $N$. In particular, 
we consider Alltop sequences for non-orthogonal
preambles as an example, while different sequences can also be used
(e.g., Zadoff-Chu sequences \cite{Chu72}).
With Alltop sequences, we have $L = N^2$ for a prime $N \ge 5$
\cite{Foucart13}.

Let $\bx_l$ denote the $l$th Alltop sequence of length $N$
with $||\bx_l|| = 1$ for all $l$.
Denote by $L_i$ the number of preamble sequences
assigned to type-$i$ devices.
Let $\cL_1 = \{\bx_1, \ldots, \bx_N\}$ 
be the set of preambles for type-1 devices with $L_1 = N$.
In addition, 
let $\cL_2 = \{\bx_{N+1}, \ldots, \bx_{N+L_2}\}$
denote the set of preambles for type-2 devices
with $L_2 \le N (N-1)$.
For convenience, the $l$th preambles of
$\cL_1$ and $\cL_2$ are denoted by $\bc_l$
and $\bar \bc_l$, respectively.
Throughout the paper, we assume that $\cL_1$ is a
set of orthogonal preambles. On the other hand,
$\cL_2$ is a set of non-orthogonal preambles.
Since Alltop sequences are used,
the correlation between any two non-orthogonal
preambles is $\frac{1}{\sqrt{N}}$. Thus,
the correlation
between any two different preambles in $\cL_2$
is $\frac{1}{\sqrt{N}}$,
while that in $\cL_1$ is 0.
The reason why orthogonal sequences are used for $\cL_1$
will be explained later.
In addition, the correlation 
between one in $\cL_1$ and another one in $\cL_2$
is $\frac{1}{\sqrt{N}}$.
It is noteworthy that although $L_2$ can be as large
as $N (N-1)$ and a large $L_2$ can lower the preamble
collision for type-2 devices,
a large $L_2$ may not be desirable 
in terms of the complexity and performance of the preamble detection 
at the BS (this issue will be discussed in detail
in Sections~\ref{S:PD2} and ~\ref{S:Sim}).

\subsection{Power-domain NOMA}

Denote by $\bh_k$ and $\bar \bh_k$ 
the channel vector of 
the $k$th active device
of type-1 and type-2, respectively.
Let $M$ denote the number of antennas at the BS. Thus,
$\bh_k, \bar \bh_k \in \uC^{M \times 1}$.
For power-domain NOMA with two different pools of preambles, 
we consider the following assumption based on \cite{Bjornson16}.
\begin{itemize}
\item[{\bf A1})]
Let $P_{{\rm tx}, k}$ and $\bar P_{{\rm tx}, k}$ 
denote the transmit powers of the the $k$th active 
type-1 and type-2 devices, respectively.
Then, 
$P_{{\rm tx},k}$ and $\bar P_{{\rm tx},k}$
are decided to be inversely proportional to
the distance between the BS and the active devices
{\color{blue} to compensate path loss}
via power control so that
\begin{align}
\bh_k \sqrt{P_{{\rm tx},k}} & = \bv_k \sqrt{P_1} \cr
\bar \bh_k \sqrt{\bar P_{{\rm tx},k}} & = \bar \bv_k \sqrt{P_2},
        \label{EQ:A}
\end{align}
where $P_i$ represents the (average) receive signal power
for type-$i$ devices, $i \in \{1,2\}$
and $\bv_k, \bar \bv_k \sim \cC \cN(\b0, \bI)$ are independent for all
$k$ (i.e., Rayleigh fading is assumed for small-scale fading).
\end{itemize}

\begin{figure}[thb]
\begin{center}
\includegraphics[width=\figwidth]{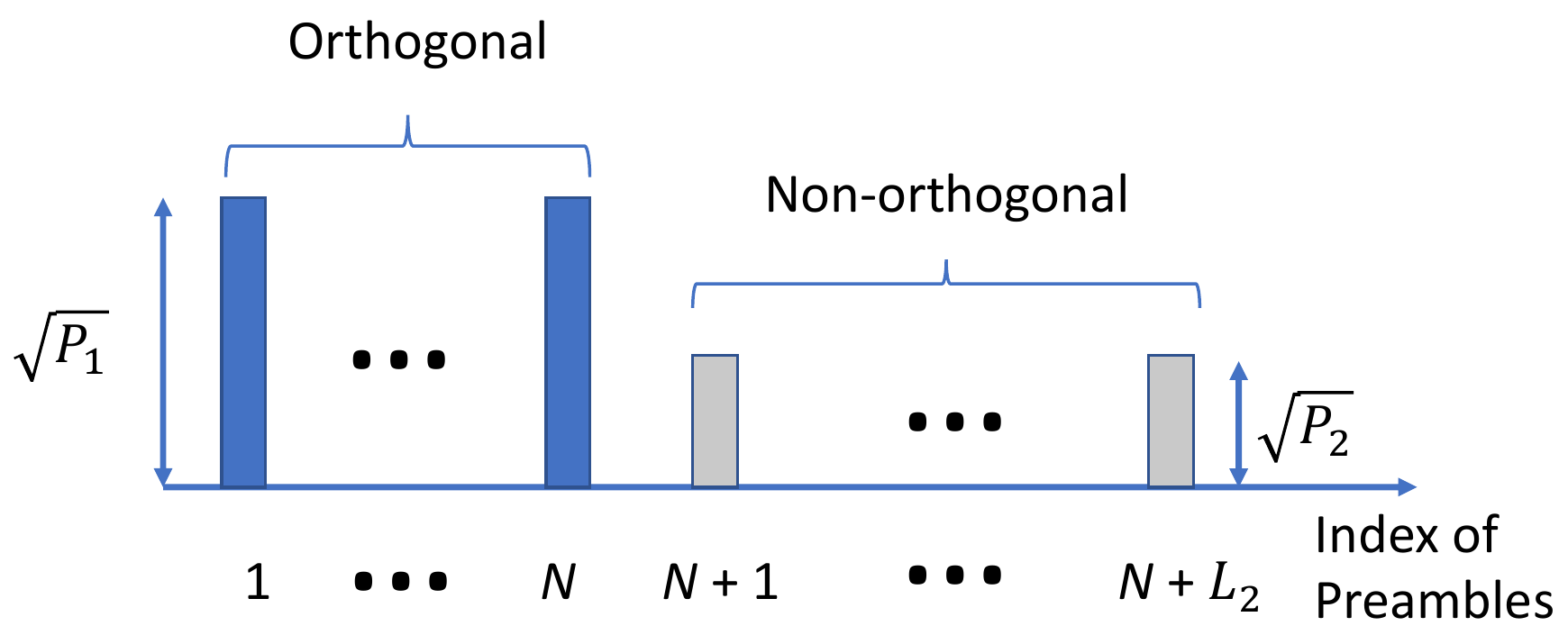}
\end{center}
\caption{Two sets of preambles for type-1 and type-2 devices with different
power levels for RALP.}
        \label{Fig:Fig1}
\end{figure}

To ensure different priorities, we assume that $P_1 > P_2$. 
Thus, as shown in Fig.~\ref{Fig:Fig1},
the preambles for type-1 devices (i.e., $\cL_1$)
are not only orthogonal, but also transmitted with a higher power
than those for type-2 devices (i.e., $\cL_2$).
As a result, type-1 devices' preambles can be more
reliably detected than type-2 devices' preambles
\emph{without any interference between active type-1 devices
of high receive power, $P_1$}. For convenience,
the resulting approach to random access with priority
is referred to as RALP.
The preamble detection for RALP
will be studied in Sections~\ref{S:PD1}
and \ref{S:PD2}.

Denote by $\cK_{i,l}$ the index set of the active type-$i$ devices
that choose preamble $l$ in $\cL_i$.
Thus, the index set of all active type-$i$ devices, denoted by
$\cK_i$, becomes $\cK_i = \cup_l \cK_{i,l}$. Let $K_i$
be the number of active type-$i$ devices,
i.e., $K_i = |\cK_i|$. Note that since
each active device chooses only one preamble,
$K_i = \sum_l K_{i,l}$, where $K_{i,l} = |\cK_{i,l}|$.
Let
\begin{align}
\bs_l & = \sum_{k \in \cK_{1,l}}
\bv_k , \ l = 1, \ldots, L_1, \cr
\bar \bs_l & = \sum_{k \in \cK_{2,l}}
\bar \bv_k  , \ l = 1, \ldots, L_2,
	\label{EQ:ss}
\end{align}
to represent the superposition of the channel vectors
associated with the active devices that choose the same preamble.
Then, the received signal at the BS 
is given by
\be
\bY = \sum_{l=1}^{L_1} \sqrt{P_1} \bs_l \bc_l^\rH
+ \sum_{l=1}^{L_2} \sqrt{P_2} \bar \bs_l  \bar \bc_l^\rH + \bN
\in \uC^{M \times N},
	\label{EQ:Y}
\ee
where $[\bN]_{m,n} \sim \cC \cN(0, N_0)$
represents the background noise.
In \eqref{EQ:Y}, the $m$th row of $\bY$ 
represents the received signal at the $m$th antenna
when active devices transmit randomly selected preambles.

As shown in \eqref{EQ:Y},
since $P_1 > P_2$, the received signals from active type-1 devices
are stronger than those from active type-2 devices. As a result,
the BS may need to detect the preambles transmitted from
active type-1 devices first. Once they are detected,
they can be removed (or suppressed) for the detection
of the preambles transmitted from
active type-2 devices.
We discuss low-complexity detection approaches in Sections~\ref{S:PD1}
and \ref{S:PD2}.

\section{Detection of Preambles Transmitted by Type-1 Devices}
\label{S:PD1}

In general, for the detection of transmitted
preambles, which is also called the user activity detection
\cite{Zhu11} \cite{Applebaum12},
there are optimal approaches based on joint detection.
%where $\bc_l$ and $\bar \bc_l$ are jointly detection.
In this case, the complexity
is proportional to $2^{L_1} \times 2^{L_2}
= 2^{L_1 + L_2}$, which is prohibitively high. As a result, 
we may resort to low-complexity suboptimal detection
approaches. To this end, we consider a two-step approach for RALP. 
In the first step, the detection of 
preambles transmitted by type-1 devices, which 
is referred to as the type-1 preamble detection,
is carried out by taking advantage of the
orthogonality of their preambles (i.e., $\cL_1$) and $P_1 > P_2$.
In the second step, all the 
preambles transmitted by type-1 devices are removed
and the detection of 
the preambles transmitted by type-2 devices
(which is also referred to as the type-2
preamble detection)
is carried out.
In this section, we focus on the first step 
and analyze the performance of preamble detection
in terms of $P_1$ and $P_2$.

\subsection{Correlator Detector}

%Since the receive power of type-1 devices
%is higher than that of type-2 devices, 
%the BS needs to detect transmitted preambles
%of active type-1 devices.
Taking advantage of the orthogonality of
$\cL_1$ (i.e., the preambles for type-1 devices),
the BS can detect them using the following correlator' output:
\begin{align}
\bg_l & = \bY \bc_l  \cr
& = \sqrt{P_1} \bs_l + 
\sqrt{P_2}
\sum_{t=1}^{L_2} \bar \bs_t 
\bar \bc_t^\rH \bc_l + \bn_l, \ l = 1, \ldots, L_1,
	\label{EQ:bgl}
\end{align}
where $\bn_l = \bN \bc_l$.
Clearly, due to the orthogonality of $\cL_1$,
there is no interference from the other active type-1 devices
of high receive power. 

Letting $\bar \bc_t^\rH \bc_l = \rho_{l,t}$,
the $m$th element of $\bg_l$ corresponding to the $m$th antenna is given by
\be
g_{m,l} = \sqrt{P_1} s_{m,l} +
\sum_{t= 1}^{L_2} \rho_{l,t} \sqrt{P_2} \bar s_{m,t} + n_{m,l},
\ee
where 
$g_{m,l}$, $s_{m,l}$, $\bar s_{m,l}$, and $n_{m,l}$ are the $m$th elements
of 
$\bg_{l}$, $\bs_{l}$, $\bar 
\bs_{l}$, and $\bn_{l}$, respectively.
Then, according to the assumption of {\bf A1}), since each element
of channel vectors is independent CSCG
and $|\rho_{l,t}|= \frac{1}{\sqrt{N}}$,
it can be seen that
\be
\sum_{t= 1}^{L_2} \rho_{l,t} \sqrt{P_2} \bar s_{m,t} + n_{m,l}
\sim \cC \cN \left(0, I_2 \right),
	\label{EQ:IG1}
\ee
where $I_2 = \frac{K_2 P_2}{N} + N_0$.
As a result, the detection of $\bc_l$ in $\bY$
can be carried out with the correlator's output, $\bg_l$
in \eqref{EQ:bgl},
which can be seen as Gaussian signal detection
in the presence of Gaussian noise that is in
\eqref{EQ:IG1}.

\subsection{Hypothesis Testing and Performance Analysis}

For two hypotheses,
letting $\cH_0$ and $\cH_1$ denote the cases of 
$K_{1,l} = 0$ (i.e., there is no type-1 device
that chooses $\bc_l$) and 
$K_{1,l} = 1$ (i.e., there is only one type-1 device
that chooses $\bc_l$), respectively, 
we have
\begin{align*}
{\cH_0\!:} \ g_{m,l} \sim \cC \cN(0,  I_2) \ \mbox{versus} \
{\cH_1\!:} \ g_{m,l} \sim \cC \cN(0, P_1 + I_2),
\end{align*}
which is Gaussian signal detection as mentioned earlier.
In addition, since
$|g_{m,l}|^2$ follows an exponential distribution,
the test statistic, $Z = ||\bg_l||^2$, follows a Gamma 
distribution and the following hypothesis testing
including the hypothesis that there are multiple 
active type-1 devices
choosing $\bc_l$, denoted by
$\cH_{\rm c}$, can be formulated:
\begin{align}
\cH_0\!: & \ Z \sim f_0 (z) = {\rm Gamma}(M, I_2)  \cr
\cH_1\!: & \ Z \sim f_1 (z) = {\rm Gamma} (M, P_1 + I_2) \cr
\cH_{\rm c}\!: & \ Z \sim f_{\rm c} (z) = {\rm Gamma} (M, K_{1,l} P_1 + I_2),
K_{1,l} \ge 2,
	\label{EQ:ZsG}
\end{align}
where ${\rm Gamma}(n, \theta) = \frac{x^{n-1} e^{-\frac{x}{\theta}}}
{\theta^n \Gamma(n)}$, for $x \ge 0$
with $n, \theta > 0$, is the Gamma distribution and
$\Gamma(n)$ is the Gamma function.

According to \eqref{EQ:ZsG},
there can be two decision threshold values, $\tau_1$ and $\tau_2$,
and decision can be carried out as follows:
\begin{align}
Z \le \tau_1:& \ \mbox{Accept $\cH_0$} \cr
\tau_1 < Z \le \tau_2:& \ \mbox{Accept $\cH_1$} \cr
Z > \tau_2:& \ \mbox{Accept $\cH_{\rm c}$}.
\end{align}
Furthermore, it is also possible to determine parameters
(e.g., $P_i$ and $\tau_i$) for  a certain target performance.
Denote by $\uP_{d|f}$ the error probability when
$\cH_d$, $d \in \{0,1,{\rm c}\}$, accepted, when $\cH_f$ is true.
The probabilities of missed detection (MD) and 
false alarm (FA) 
are given by
\begin{align}
\uP_{0, {\rm MD}} & = \uP_{0|1} = \int_0^{\tau_1} f_1 (z) d z \cr
\uP_{\rm c, MD} & = \uP_{{\rm c}|1}  
= \int_{\tau_2}^\infty f_1 (z) d z \cr
\uP_{1, {\rm MD}} & = \uP_{1| {\rm c}} 
= \int_{\tau_1}^{\tau_2} f_{\rm c} (z) d z \cr
\uP_{\rm FA} & = \uP_{1|0} + \uP_{{\rm c}|0}
= \int_{\tau_1}^\infty f_0 (z) d z.
	\label{EQ:Pes}
\end{align}
Using the Gamma distributions in \eqref{EQ:ZsG},
all the error probabilities in
\eqref{EQ:Pes} can be found as closed-form expressions.
We have a few remarks on the error events.

\begin{itemize}

\item
There are three events of MD associated with the first
three probabilities in \eqref{EQ:Pes}.
The events of MD when $\cH_1$ is true may lead to re-transmissions
by the related active type-1 devices,
while the event of MD when $\cH_{\rm c}$ is true
has to be rectified by further steps in the handshaking process.
In addition, with layered preambles, 
the event of MD associated with $\uP_{0|1}$ 
leads to error propagation and high interference
that degrades the performance of preamble detection of
type-2 devices, while that with $\uP_{{\rm c}|1}$ 
leads to the signal dimension reduction has less serious impact
on the performance as will be discussed 
in Section~\ref{S:PD2}.

\item In general, an event of FA results in 
incorrect acknowledgment of successful preamble transmission
to an inactive device in the handshaking process, which is
disregarded by the inactive device. 
Thus, it might be tolerable to have a relatively high
probability of FA.
However, with the proposed layered preambles,
any FA events can lead to error propagation through 
SIC
and a degraded performance of type-2 preamble detection.
However, the resulting performance
degradation is not serious as that due to the event of MD
associated with $\uP_{0|1}$, 
which will be explained in Section~\ref{S:PD2}.
\end{itemize}

A salient feature of the proposed approach, i.e., RALP, 
is that it can have guaranteed performance for type-1 devices,
because the error probabilities can be found 
with given parameters as shown in \eqref{EQ:Pes}
(thanks to the simple detection approach (i.e., the correlator
detector) whose performance can be simply 
obtained as closed-form expressions\footnote{On the other hand,
the performance of type-2 preamble detection
is not easily obtained as closed-form expressions.}
in \eqref{EQ:Pes}).

\subsection{Key Error Probabilities}

In this subsection, we study key error probabilities
of the type-1 preamble detection.

Provided that $K_1 \ll L_1 = N$ and $P_1$ is sufficiently high,
it is expected that the events of MD associated with
$\uP_{{\rm c}| 1}$ and $\uP_{1|{\rm c}}$ in \eqref{EQ:Pes} may not 
frequently happen.
Thus, we may focus on the first two probabilities of errors,
i.e., $\uP_{\rm FA}$ and $\uP_{\rm MD} = \uP_{0|1}$,
in \eqref{EQ:Pes},
which is mainly decided by $\tau_1$.
With a small $\epsilon > 0$, $\tau_1$ can be decided to keep
\be
\uP_{\rm MD} \le \epsilon
	\label{EQ:Peps}
\ee
to not only minimize the impact of error propagation 
and high interference
on the performance of type-2 preamble detection,
but also reduce the number of re-transmissions 
for low access delay.

Note that although $\epsilon \to 0$,
there are active type-1 devices that fail to transmit their
preambles because of preamble collisions.
Thus, with a sufficiently low $\epsilon$,
re-transmissions are mainly caused by 
preamble collisions for type-1 devices.
As a result, for a low access delay, 
the probability of preamble collision 
has to be controlled by limiting
the number of type-1 devices per RB such that $\uE[K_1] \ll N$.
This issue is related user barring \cite{Li17_S},
which is beyond the scope of the paper.

From \eqref{EQ:ZsG}, \eqref{EQ:Peps}
can be rewritten as 
\be
\uP_{\rm MD} =
F_1 (\tau_1)
= e^{-\nu_1}\sum_{m=M}^\infty \frac{ \nu_1^m}{m!}
\le \epsilon,
	\label{EQ:uFe}
\ee
where $F_i (\cdot)$ represents
the cdf of $f_i (\cdot)$ and
$\nu_1 =  \frac{\tau_1}{P_1 + I_2}$.
Clearly, $F_1 (\tau_1)$ corresponds to
the probability that a Poisson random variable
with mean $\nu_1$ is greater than or equal to $M$.
Thus, for a low $\uP_{\rm MD}$, $\nu_1 < M$ is required.

\begin{figure}[thb]
\begin{center}
\includegraphics[width=\figwidth]{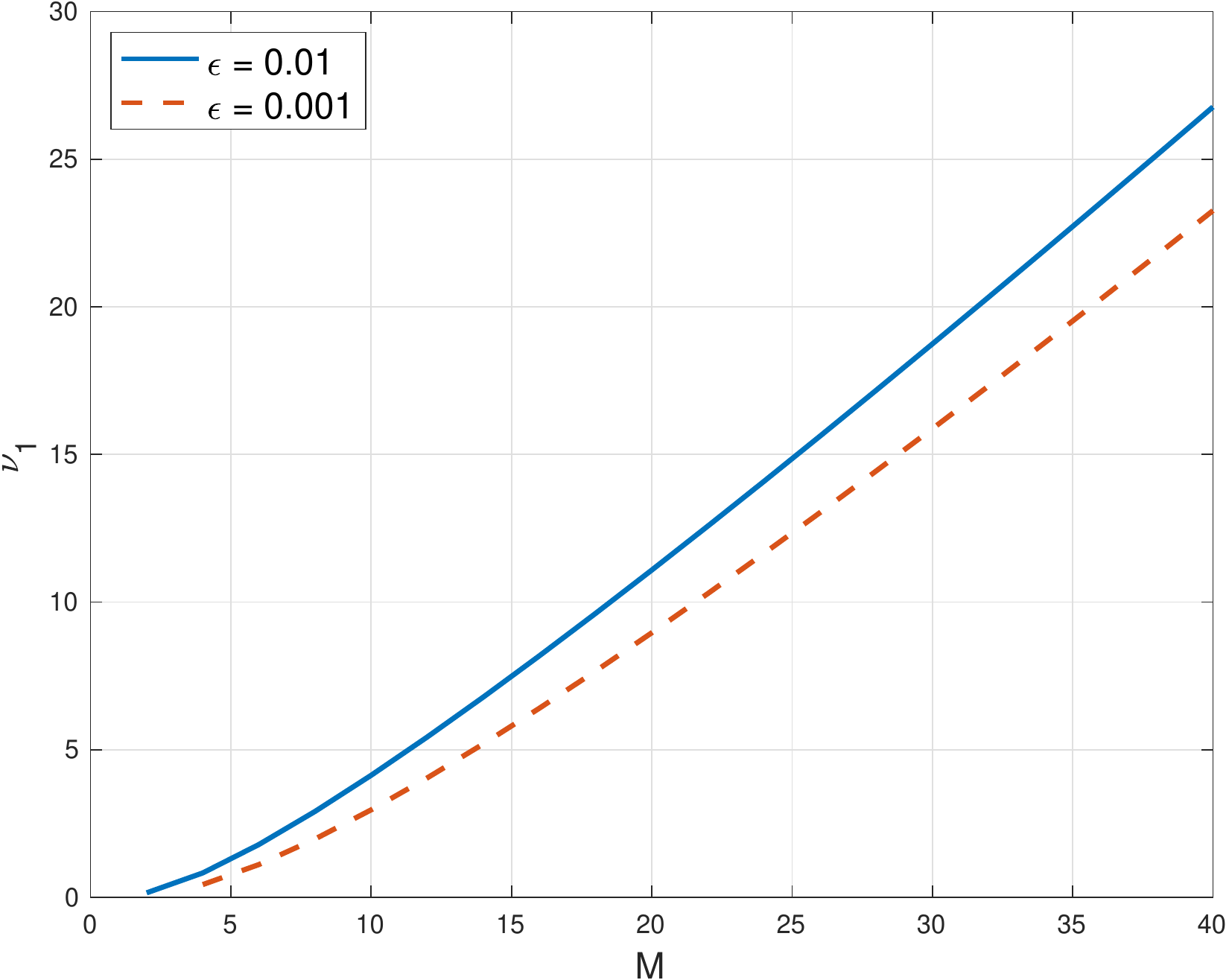}
\end{center}
\caption{Two sets of preambles for type-1 and type-2 devices with different
power levels.}
        \label{Fig:M_nu}
\end{figure}

Fig.~\ref{Fig:M_nu} shows
the value of $\nu_1$ as a function of $M$ for a given $\epsilon
\in \{10^{-2}, 10^{-3}\}$.
Denote by $\nu_1(M,\epsilon)$
the value of $\nu_1$ that satisfies the quality in \eqref{EQ:uFe}
for given $\epsilon$ and $M$. 
Consequently, we can have the following relationship:
\be
\nu_1(M,\epsilon) = \frac{\tau_1}{P_1 + \frac{K_2 P_2}{N} + N_0},
\ee
which can be used to decide $\tau_1$ for given $P_1$, $K_2$, and $P_2$.

\begin{figure}[thb]
\begin{center}
\includegraphics[width=\figwidth]{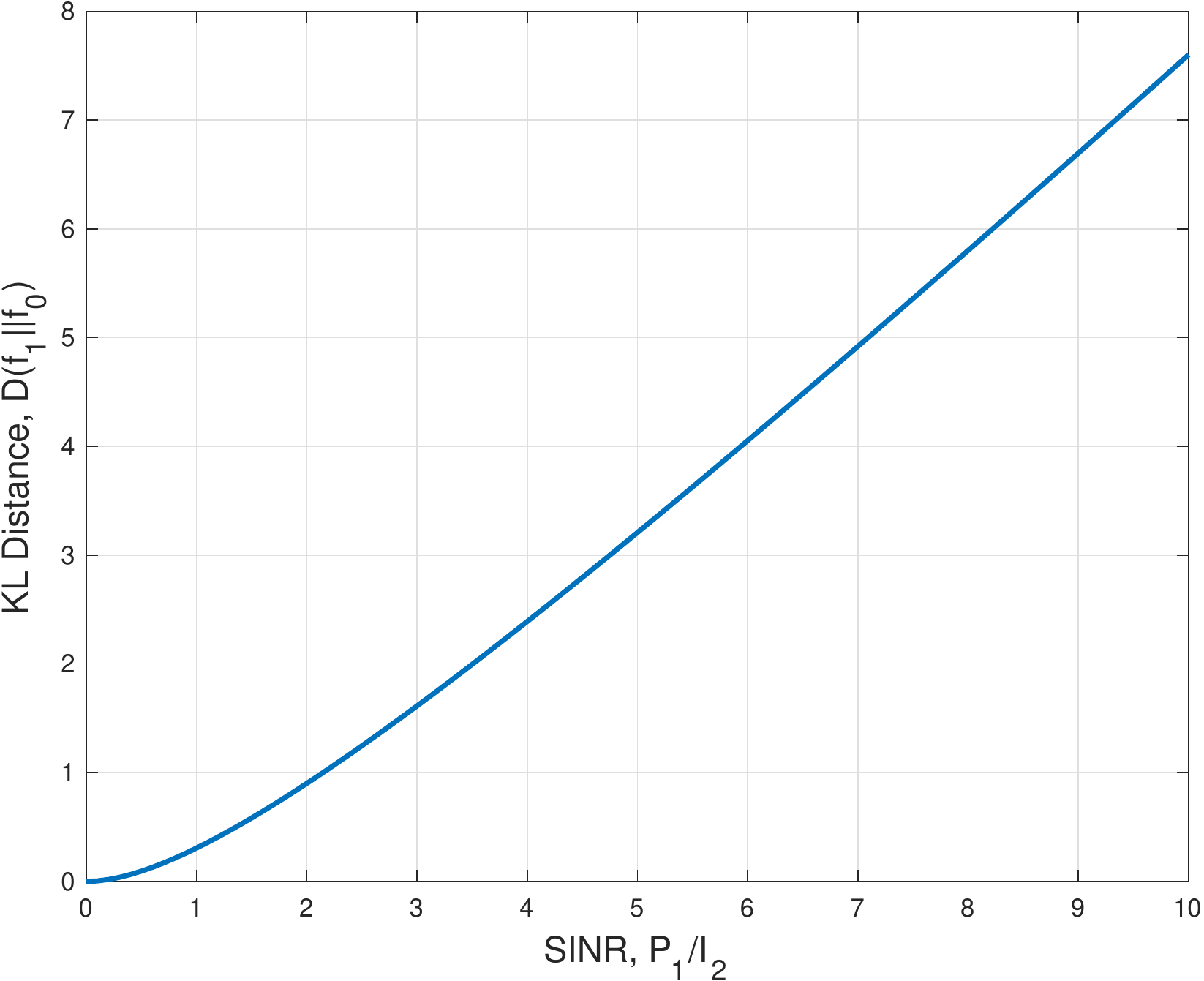}
\end{center}
\caption{The KD distance in \eqref{EQ:KLD} as a function of 
$P_1/I_2$.}
        \label{Fig:D_PI}
\end{figure}

According to \cite{CoverBook}, we have
\be
\lim_{\epsilon \to 0}
\lim_{M \to \infty}
\frac{1}{M}
\ln \uP_{\rm FA} 
= - \sD(\tilde f_1 \,||\, \tilde f_0),
	\label{EQ:eFA}
\ee
where $\sD(\tilde f_1 \,||\, \tilde f_0)
= \uE_0 \left[\ln \frac{\tilde f_1 (Z)}{\tilde f_0 (Z)} \right] $ is the 
Kullback Leibler (KL) divergence or distance when $M = 1$.
Here, 
$\tilde f_i (z)$ represents $f_i (z)$, $i = 0,1$, with $M = 1$ and
$\uE_0 [\cdot]$ represents the expectation with respect to
$\tilde f_0(z)$.
From \eqref{EQ:ZsG},
it can be shown that
\be
\sD(\tilde f_1 \,||\, \tilde f_0)
=
\frac{P_1}{I_2} - \ln \left( 1+ \frac{P_1}{I_2} \right)  \ge 0.
	\label{EQ:KLD}
\ee
Consequently, with $\uP_{\rm MD} = \uP_{0|1} \le \epsilon$,
we can see that $\uP_{\rm FA}$ decreases exponentially with $M$
(i.e., a large $M$ can result in a low 
$\uP_{\rm FA}$).
In Fig.~\ref{Fig:D_PI},
the KL distance in \eqref{EQ:KLD}
is shown as a function of $P_1/I_2$.

\section{Detection of Preambles Transmitted by Type-2 Devices}	\label{S:PD2}

In this section, we discuss the type-2 preamble detection
based on the notion of sparse signal recovery
under the assumption that $K_2 \ll L_2$.

\subsection{SIC and Error Propagation}

Provided that the type-1 preamble detection 
is successfully carried out, the type-2 preamble detection
can be considered with SIC.
For convenience, let $\cJ_i$ denote the index set of
the preambles that are chosen by active type-$i$ devices,
i.e.,
$$
\cJ_i = \{l: \ K_{i,l}  \ge 1\}.
$$
For $l \in \cJ_1$, we assume that
$\frac{1}{\sqrt{P_1}} \bg_l$ is an estimate of $\bs_l$.
Thus, for SIC, the received signal from active type-1 devices can be
reconstructed and removed as follows:
\begin{align}
\bar \bY & = \bY - \sum_{l \in \cJ_1}  \bg_l \bc_l^\rH =  \bY \bP_1 \cr
& = \left(\sum_{l=1}^{L_2} \sqrt{P_2} \bar \bs_l
\bar \bc_l^\rH + \bN \right) \bP_1,
	\label{EQ:bY}
\end{align}
where $\bP_1 = \bI - \sum_{l \in \cJ_1} \bc_l \bc_l^\rH$ is an
orthogonal projection matrix.
This implies that SIC results in the signal suppression
that suppresses all the signals in the subspace
spanned by $\bc_l$, $l \in \cJ_1$.

From \eqref{EQ:bY}, we can see 
the impact of error propagation on the performance
of type-2 preamble detection.
For the event of FA, suppose that $\bc_1$ in $\cL_1$ is incorrectly
detected as a transmitted one.
In this case, the orthogonal projection matrix becomes
\be
\bP_1 = \bI - \sum_{l \in \cJ_1} \bc_l \bc_l^\rH
- \bc_1 \bc_1^\rH.
\ee
This results in unnecessary signal suppression 
associated with the subspace spanned by $\bc_1$,
which leads to the dimension reduction.

For the event of MD, suppose that $\bc_1$ belongs
to $\cJ_1$ (i.e., it is a transmitted preamble), but
it is not detected. 
Then, $\bP_1$ is modified as
$\bP_1 = \bI - \sum_{l \in \cJ_1} \bc_l \bc_l^\rH
+ \bc_1 \bc_1^\rH$,
which leads to 
\be
\bar \bY = \sqrt{P_1} \bs_1 \bc_1^\rH +
\left(\sum_{l=1}^{L_2} \sqrt{P_2} \bar \bs_l
\bar \bc_l^\rH + \bN \right) \bP_1.
\ee
Clearly, the type-2 preamble detection
suffers from the strong interference due to
undetected and unsuppressed preambles of type-1 devices.
Compared to the dimension reduction due to FA,
the presence of strong interference due to MD may result in a 
severe performance degradation,
which will be demonstrated in Section~\ref{S:Sim}.

\subsection{Preamble Detection as Sparse Vector Estimation}

In this subsection, we consider the type-2 preamble detection
after SIC as a sparse vector estimation
problem. For simplicity, it is assumed that there are no
FA and MD errors when detecting preambles from type-1 devices.

Let $\bC_{\rm u}$ be the matrix consisting of the column
vectors that are $\bc_l$, $l \notin \cJ_1$.
Clearly, $\bC_{\rm u}$ is orthogonal to $\bc_l$, $l \in \cJ_1$
and $\bP_1 \bC_{\rm u} = \bC_{\rm u}$.
Denote by $\bar \br_m^\rH $ the $m$th row of $\bar \bY$ and
let $\bz_m = 
(\bar \br_m^\rH \bC_{\rm u})^\rH =
\bC_{\rm u}^\rH \bar \br_m$ and $\bar \bC = [\bar \bc_1 
\ \ldots \ \bar \bc_{L_2}]$.
Then, after some manipulations,
it can be shown that
\begin{align}
\bz_m = \underbrace{\bC_{\rm u}^\rH \bar \bC}_{=
\bPhi} \ba_m + \bu_m, \ m = 1, \ldots, M,
	\label{EQ:bz_m}
\end{align}
where $\bu_m^\rH$ is the $m$th row
of $\bN \bP_1 \bC_{\rm u}$, $\bu_m \sim \cC \cN \left(\b0, N_0 \bI \right)$, 
and
\be
[\ba_m]_l = \sqrt{P_2} \bar s_{m,l}^*.
	\label{EQ:bam}
\ee
Since there are $K_2$ active type-2 devices,
we can see that
\be
||\ba_m||_0 \le K_2 \ \mbox{and} \ 
{\rm supp}(\ba_m) = {\rm supp}(\ba_{m^\prime}), \ m, m^\prime = 1,\ldots,M.
\ee
In addition, letting $J_1 = |\cJ_1|$. $\bPhi$ becomes
a $J_1 \times L_2$ matrix.
Let $\bU = [\bu_1 \ \ldots \ \bu_M] \in \uC^{J_1 \times M}$
and $\bZ = [\bz_1 \ \ldots \ \bz_M] \in \uC^{J_1 \times M}$.
Then, it can be shown that
\be
\bZ = \bPhi \bA + \bU,
	\label{EQ:MMV}
\ee
where $\bA = [\ba_1 \ \ldots \ \ba_M]$ is a row-sparse matrix
\cite{Chen06} \cite{Davies12} \cite{Eldar12}.
As a result, finding the non-zero rows which is
equivalent to the detection of transmitted preambles
by type-2 devices is a multiple measurement vectors (MMV) problem.

There have been a number of approaches and algorithms proposed to
solve MMV problems. In general, their complexity
increases with the number of columns of $\bPhi$, $L_2$.
In addition, for a fixed row-sparsity, 
a better performance is achieved with a smaller number of columns, $L_2$.
Thus, although a large $L_2$ is desirable for a low probability
of preamble collision, it is necessary to keep $L_2$ as small 
as possible so that the complexity of algorithms
is sufficiently low with good performance.

\subsection{CAVI Algorithm for Low-Complexity Detection} \label{SS:CAVI}

There are a number of approaches to MMV problems.
Among those, 
to detect transmitted preambles by type-2 devices in this section,
we consider an approach
based on variational inference,
namely the coordinate ascent VI 
(CAVI) algorithm
\cite{Bishop06} \cite{Blei17}, which has been successfully
used in \cite{Choi19a} to detect sparse signals in MTC.

Let $\bb$ denote the binary vector of length $L_2$,
where $b_l = 1$ if $\bar \bc_l$ is transmitted and $0$
otherwise. Here, $b_l$ is referred to as the activity variable.
Then, $\ba_m$ in \eqref{EQ:bam} can be represented 
as 
\be
\ba_m = \sqrt{P_2} \bW_m \bb,
	\label{EQ:aPWb}
\ee
where $\bW_m$ is a diagonal matrix.
To see the elements of $\bW_m$, consider
the variance of $\bar s_{m,l}^*$.
For $l \in \cJ_2$, $\bar s_{m,l}^*$ is a zero-mean
CSCG random variable according to the assumption of {\bf A1})
and \eqref{EQ:ss}.
In addition, its variance becomes $1$ if only one type-2 device
chooses $\bar \bc_l$. If more devices choose $\bar \bc_l$,
the variance increases. Thus, we have
\be
{\rm Var} (\bar s_{m,l}) = 
\uE[ K_{2,l} \,|\, K_{2,l} \ge 1], \ l \in \cJ_2,
\ee
which is denoted by $\sigma_s^2$.  
Let
\be
[\bW_m]_{l,l}
= \left\{
\begin{array}{ll}
\sqrt{P_2} \bar s_{m,l}^* \sim \cC \cN(0, P_2 \sigma_s^2), 
& \mbox{if $l \in \cJ_2$} \cr
\cC \cN(0, P_2 \sigma_s^2), & \mbox{o.w.}
\end{array}
\right.
	\label{EQ:Wm}
\ee
so that \eqref{EQ:aPWb} is valid.
Clearly, the diagonal elements of $\bW_m$ 
in \eqref{EQ:Wm} are iid and CSCG.
Then, $\bz_m$ in \eqref{EQ:bz_m}
can also be expressed as
\be
\bz_m = \bPhi \bW_m \bb + \bu_m,
\ee
which can be characterized as the following CSCG random vector:
\be
\bz_m \sim \cC \cN(\b0, P_2 \sigma_s^2 \bPhi 
{\rm diag}(\bb) \bPhi^\rH + N_0 \bI), \ m = 1,\ldots, M.
	\label{EQ:z_CN}
\ee
Note that in \eqref{EQ:z_CN}, $\sigma_s^2$ is to be decided.
Suppose that the number of active type-2 devices, $K_2$,
follows a Poisson distribution with mean
$\lambda_2$. Since each active type-2 device chooses one of
$L_2$ preambles in $\cL_2$ uniformly at random,
$K_{2,l}$ becomes a Poisson random variable with 
mean $\frac{\lambda_2}{L_2}$, i.e., 
$K_{2,l} \sim {\rm Pois} \left(
\frac{\lambda_2}{L_2} \right)$.
Then, it can be shown that
\begin{align}
\uE[ K_{2,l} \,|\, K_{2,l} \ge 1]
& = \frac{\sum_{k=1}^\infty k \frac{(\lambda_2/L_2)^k}{k!} e^{-
\frac{\lambda_2}{L_2}}} {\Pr(K_{2,l} \ge 1)} \cr
& = \frac{\lambda_2/L_2}{1 - e^{-\lambda_2/L_2}}.
	\label{EQ:EKK}
\end{align}
If $\frac{\lambda_2}{L_2} \to 0$, we can see that
$\uE[ K_{2,l} \,|\, K_{2,l} \ge 1] \to 1$\footnote{Since
$\uE[ K_{2,l} \,|\, K_{2,l} \ge 1]$ 
in \eqref{EQ:EKK}
increases with $\frac{\lambda_2}{L_2}
(\ge 0)$, its minimum is 1.}.
That is, for a sufficiently small
$\frac{\lambda_2}{L_2} \to 0$, $\sigma_s^2$ can be set to 1.

Consequently, the detection of the transmitted
preambles by type-2 devices can be carried out
by the following maximum
a posteriori probability (MAP) approach \cite{ChoiJBook2}:
\begin{align}
\hat \bb & = \argmax_{\bb \in \cB} \Pr(\bb\,|\, \{\bz_m\}) \cr
& = \argmax_{\bb \in \cB} \prod_m f(\bz_m\,|\,\bb) + \Pr(\bb),
	\label{EQ:map}
\end{align}
where $\cB = \{\bb\,|\, [\bb]_l \in \{0,1\}\}$.
In \eqref{EQ:map}, the activity variables, which are binary 
random variables, are to be detected. 
If an exhaustive search is considered, the complexity
is proportional to $|\cB| = 2^{L_2}$.
To avoid this, we can consider the variational
distribution for each $b_{l}$, denoted by $\psi_l(b_l)$,
and solve the following optimization problem:
\be
\psi^* (\bb) = \argmin_{\psi(\bb) \in \Psi}
\sD (\psi (\bb) || \Pr(\bb\,|\, \{\bz_m\})),
	\label{EQ:VI}
\ee
where $\psi(\bb) = \prod_{m} \psi(b_m)$ and
$\Psi$ represents the collection of all the possible distributions of 
$\bb$. Here, the KL divergence is given by
$$
\sD (\psi(\bb)||f(\bb))
= \sum_\bb \psi(\bb) \ln \frac{\psi(\bb)}{f(\bb)},
$$
where $f(\bb)$ is any distribution of $\bb$
with $f(\bb) > 0$ for all $\bb \in \cB$.
In \eqref{EQ:VI},
clearly, we attempt to find $\psi(\bb)$ that is close
to the a posteriori probability, $\Pr(\bb\,|\, \by)$,
as an approximation.
In \cite{Blei17}, 
the minimization of 
the KL divergence is equivalent to 
the maximization of the evidence lower bound (ELBO), which is given by
$$
{\rm ELBO}(\psi) = \uE[\ln f(\{\bz_m\},\bb)] - \uE[\ln \psi(\bb)].
$$
Let $\bb_{-l} = [b_1 \ \ldots \ b_{l-1} \ b_{l+1} \ \ldots \ b_{L_2}]^\rT$
and
$\psi_{-l} (\bb_{-l}) = \sum_{i \ne l} \psi_i (b_i)$.
The CAVI algorithm \cite{Bishop06, Blei17} is to 
update one variational distribution at a time 
with the other variational distributions fixed
(so that the ELBO can be minimized through iterations) as follows:
\begin{align}
\psi_l^*(b_l) 
\propto \theta_l (b_l) \deft \exp \left(\uE_{-l} [\ln f(b_l\,|\, \bb_{-l}, 
\{\bz_m\})] \right),
	\label{EQ:theta}
\end{align}
where $\uE_{-l}[\cdot]$ 
represents the expectation with 
respect to $\bb_{-l}$ or with the distribution $\psi_{-l} (\bb_{-l})$.
Let $\psi_l^{(q)}$ denote the $q$th estimate of $\psi_l$.
In the CAVI algorithm, $\psi_l^{(q)}$ is updated from $l = 1$ to $L_2$ in each
iteration. The number of iterations is denoted by $N_{\rm run}$.
Unfortunately, the convergence behavior of the CAVI algorithm
is not known \cite{Blei17} and $N_{\rm run}$ can be decided 
through experiments \cite{Choi19a}.

Note that thanks to the Gaussian distribution (of $\{\bz_m\}$),
it is possible to find a closed-form expression
for $\theta_l (b_l)$ in \eqref{EQ:theta}, 
which can be found in \cite{Choi19a}.

There are a few remarks as follows.
\begin{itemize}
\item In general, it is not easy to find the performance
of preamble detection when preambles are not orthogonal,
which implies that the performance cannot be guaranteed 
in terms of the probabilities of MD and FA.
On the other hand, as shown in Section~\ref{S:PD1},
in RALP, at least, it is possible to guarantee certain performance
in terms of the probabilities of MD and FA
for type-1 devices thanks to their orthogonal preambles, $\cL_1$.
From this, resource allocation and barring schemes
can manage to keep QoS requirements 
for type-1 devices. 

\item It is noteworthy that the approach in \cite{Li17_S}, which
supports different priorities through dynamic allocation of
preambles, does not take into account MD events 
that incur re-transmissions like preamble collisions.
Since the probability of MD is not negligible
with non-orthogonal preambles
as will be shown in Section~\ref{S:Sim},
it is required to take into account both 
the probabilities of preamble collision and MD
so that guaranteed access delay can be ensured.
However, as mentioned earlier, with non-orthogonal preambles,
it is difficult to find the probability of MD.
To keep the probability of MD negligible,
the approach in \cite{Li17_S} 
can only be used with orthogonal preambles, which however
limits the number of devices to be supported.

\item The overall complexity of signal detection at the BS
can be divided into parts. The first is to detect
type-1 devices. Since a bank of $L_1$ correlators
is used as in \eqref{EQ:bgl},
the complexity is $O(M L_1 N^2)$.
The second is to perform the CAVI algorithm to detect
type-2 devices.
It can be shown that the complexity of the CAVI
algorithm per iteration is $O(M L_2 N^2)$ \cite{Choi19a}.
As a result, the total complexity is $O(M (L_1+ L_2) N^2)$
if the number of iterations for the CAVI algorithm
is fixed (usually 5 iterations are sufficient).
\end{itemize}

\section{Simulation Results}	\label{S:Sim}

In this section, we present simulation results
under the assumption of {\bf A1}) with Alltop
sequences of length $N \in \{13, 37\}$.
In order to focus on the probabilities of MD and FA
as performance criteria,
no preamble collisions are taken into account with fixed $K_1$
and $K_2$ (i.e., it is assumed
that each active device chooses a unique preamble
with $K_i \le L_i$).

\subsection{Performance of Type-1 Preamble Detection}

In this subsection, the performance of of type-1 
preamble detection 
is shown with the probabilities of MD and FA.
As in \eqref{EQ:Peps},
$\tau_1$ is decided to keep the probability of MD 
to be equal to or less than $\epsilon$.

\begin{figure}[thb]
\begin{center}
\includegraphics[width=\figwidth]{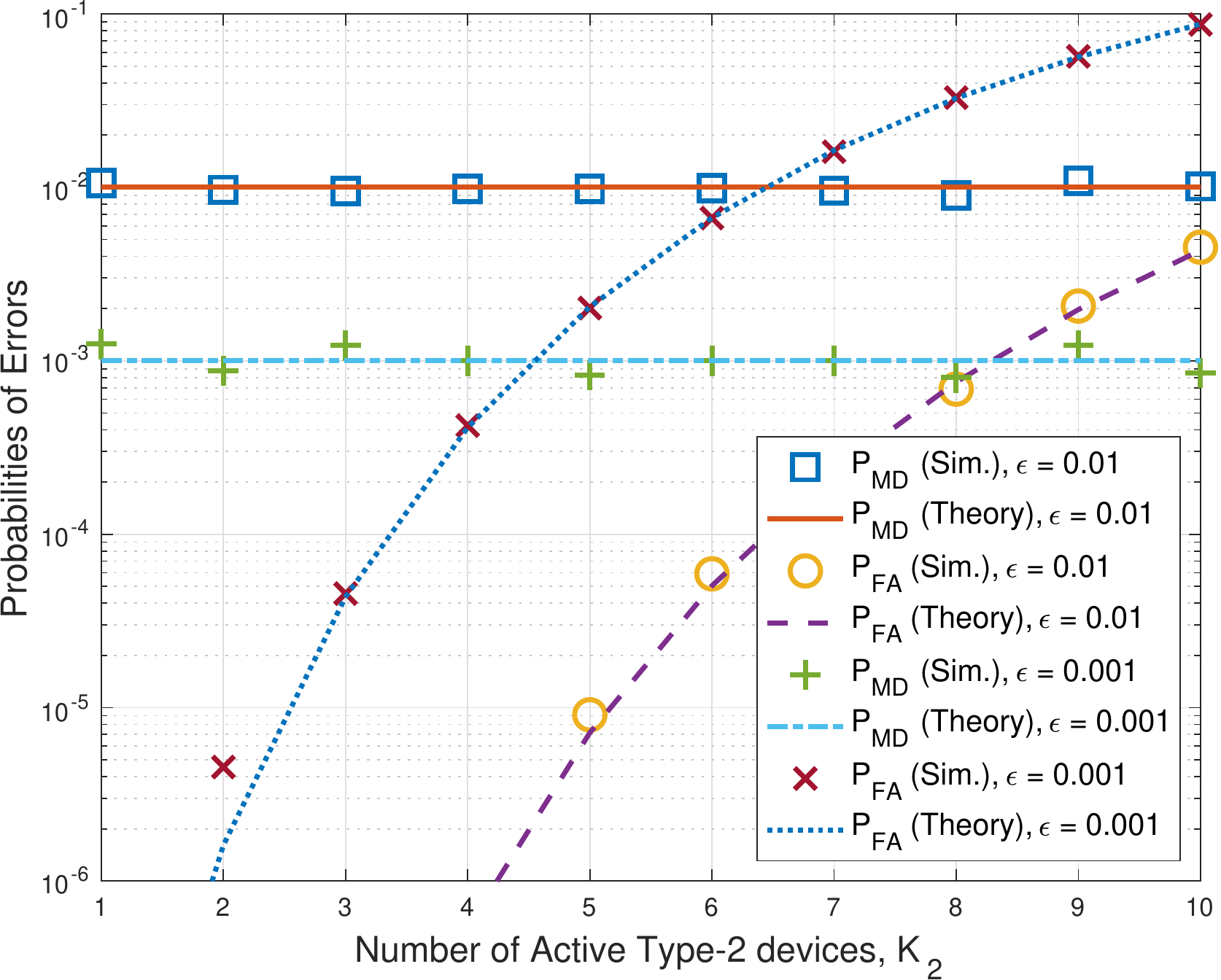}
\end{center}
\caption{The probabilities of (preamble detection) errors of 
active type-1 devices 
as functions of the number of active type-2 devices, $K_2$,
when $K_1 = 2$, $N = L_1 = 13$, $L_2 = 5N$, $M = 10$, $P_1 = 12$ dB,
$P_2 = 6$ dB, and $\epsilon \in \{10^{-2}, 10^{-3}\}$.}
        \label{Fig:t1_plt1}
\end{figure}

Fig.~\ref{Fig:t1_plt1} shows 
the probabilities of (preamble detection) errors of 
active type-1 devices 
as functions of the number of active type-2 devices, $K_2$,
when $K_1 = 2$, $N = L_1 = 13$, $L_2 = 5N$, $M = 10$, $P_1 = 12$ dB,
$P_2 = 6$ dB, and $\epsilon \in \{10^{-2}, 10^{-3}\}$.
With $\uP_{\rm MD} = \epsilon$,
it is shown that $\uP_{\rm FA}$ increases
with $K_2$ (due to the increase of the interference, $I_2$).
In addition, with a lower $\epsilon$, 
$\uP_{\rm FA}$ becomes higher.
We can also see that
the theoretical results (obtained
from \eqref{EQ:Pes}) agree with the simulation results,
which is an important observation 
as certain performance can be guaranteed for type-1 devices
by deciding key parameters using the known distribution 
of $Z = ||\bg_l||^2$ in \eqref{EQ:ZsG}.

\begin{figure}[thb]
\begin{center}
\includegraphics[width=\figwidth]{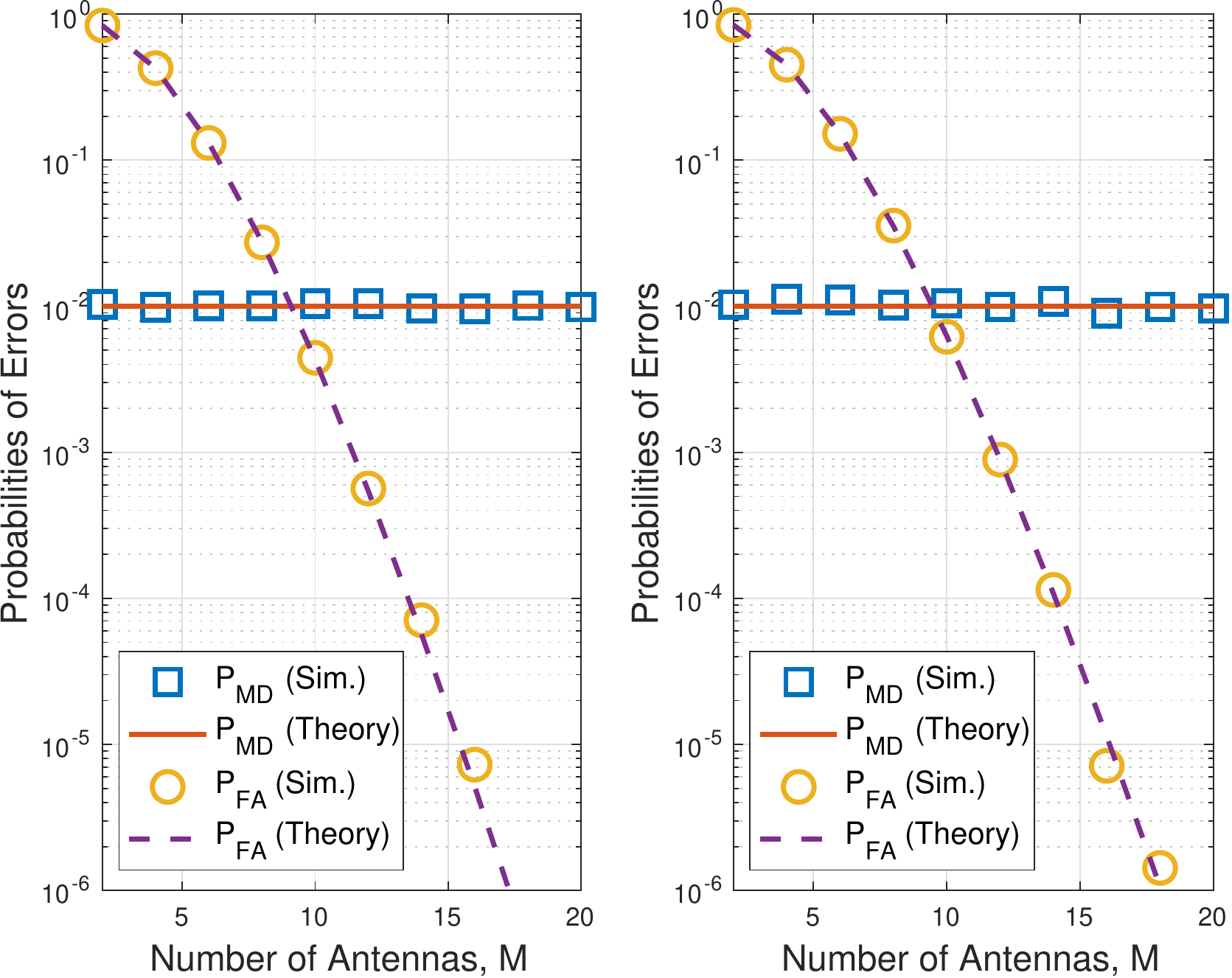} \\
\hskip 0.5cm (a) \hskip 3.5cm (b)
\end{center}
\caption{The probabilities of (preamble detection) errors of active type-1 
devices as functions of the number of antennas, $M$, at the BS
when $K_1= 2$, $L_2 = 5N$, $P_1 = 12$ dB,
$P_2 = 6$ dB, and $\epsilon = 10^{-2}$: 
(a) $N = L_1 = 13$ and $K_2 = 10$;
(b) $N = L_1 = 37$ and $K_2 = 30$.}
        \label{Fig:t1_plt2}
\end{figure}

In Fig.~\ref{Fig:t1_plt2},
the probabilities of (preamble detection) errors of active type-1 
devices 
are shown as functions of the number of antennas, $M$, at the BS 
when $K_1= 2$, $L_2 = 5N$, $P_1 = 12$ dB,
$P_2 = 6$ dB, and $\epsilon = 10^{-2}$
for small (i.e., $N = 13$) and large (i.e., $N = 37$)
systems.
Clearly, a better performance can be achieved with more antenna
elements at the BS,
which is predicted by \eqref{EQ:eFA},
i.e., the probability of FA decreases exponentially with $M$.
Note that in Figs.~\ref{Fig:t1_plt2} (a) and (b),
we have $\frac{K_2}{N} = \frac{10}{13}$ and
$\frac{K_2}{N} = \frac{30}{37}$, respectively. Thus,
$I_2$ is almost the same, which leads to almost
identical performance regardless of $N$ in
Figs.~\ref{Fig:t1_plt2} (a) and (b).

\begin{figure}[thb]
\begin{center}
\includegraphics[width=\figwidth]{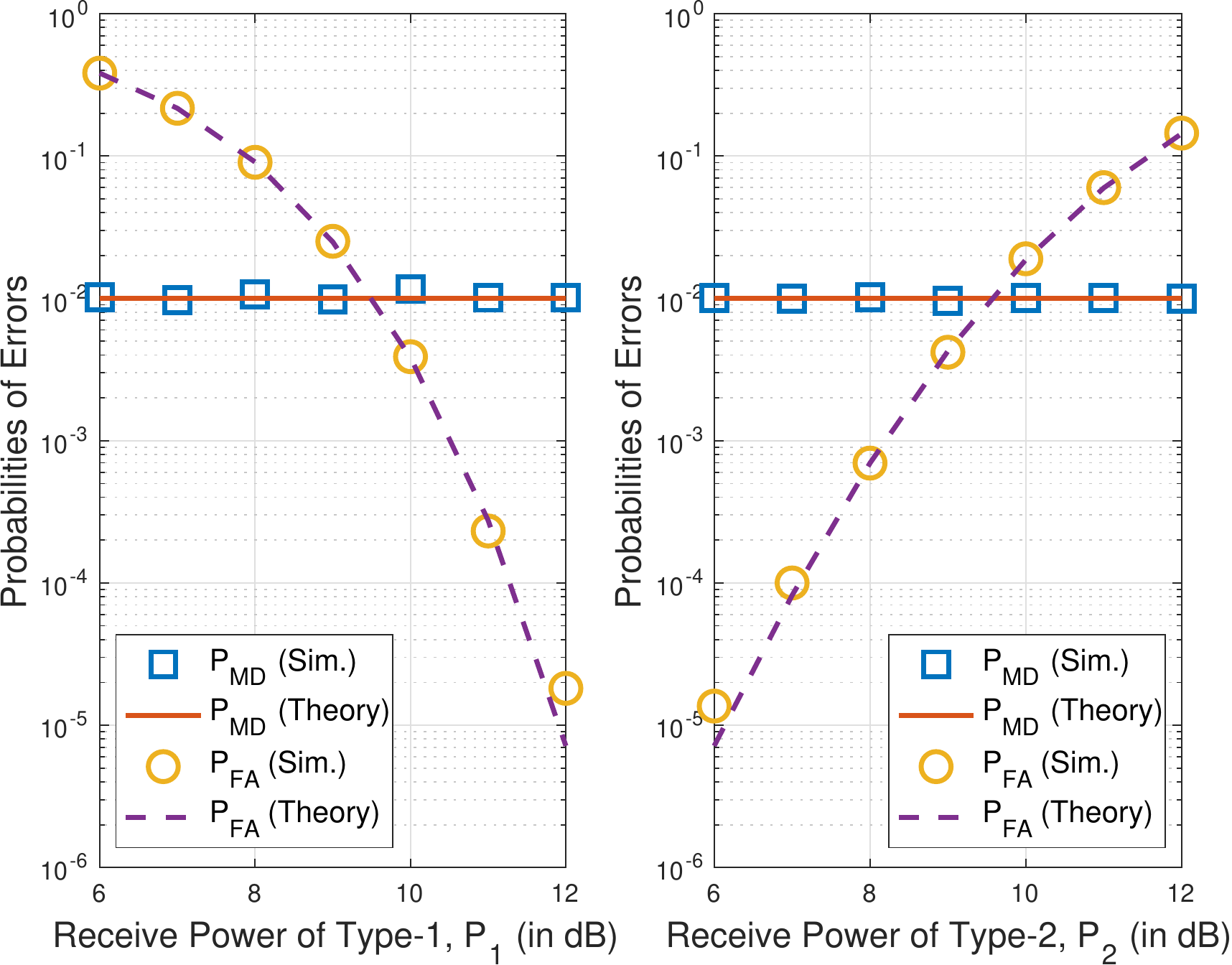} \\
\hskip 0.5cm (a) \hskip 3.5cm (b)  
\end{center}
\caption{The probabilities of (preamble detection) errors of active type-1 
devices as functions of the receive power
when $(K_1, K_2)= (2, 5)$, $N = L_1 = 13$, $L_2 = 5N$, $M = 10$,
and $\epsilon = 10^{-2}$: 
(a) Probabilities versus $P_1$ when $P_2 = 6$ dB;
(b) Probabilities versus $P_2$ when $P_1 = 12$ dB.}
        \label{Fig:t1_plt34}
\end{figure}

Fig.~\ref{Fig:t1_plt34} shows 
the probabilities of (preamble detection) errors of active type-1 
devices as functions of the receive power
when $(K_1, K_2)= (2, 5)$, $N = L_1 = 13$, $L_2 = 5N$, $M = 10$,
and $\epsilon = 10^{-2}$. 
In particular, 
in Fig.~\ref{Fig:t1_plt34} (a),
with a fixed $P_2$ (i.e., $P_2 = 6$ dB),
it is shown that the probability of FA decreases with $P_1$.
In Fig.~\ref{Fig:t1_plt34} (b),
with a fixed $P_1$ (i.e., $P_1 = 12$ dB),
due to the increase of the interference power,
the probability of FA increases with $P_2$.
Thus, with a target probability of FA,
for a given $P_2$ (or $P_1$), $P_1$ 
(or $P_2$, respectively) can be decided.

\subsection{Performance of Type-2 Preamble Detection}

In this subsection, we present simulation
results of the type-2 preamble detection 
using the CAVI algorithm in Subsection~\ref{SS:CAVI}
with $5$ iterations. 
To see the impact of error propagation 
due to MD and FA events in 
the type-1 preamble detection,
we consider two different cases: 
\emph{i)} one preamble in $\cJ_1^c$ is incorrectly
detected (i.e., an event of FA);
\emph{ii)} one preamble in $\cJ_1$ is not 
detected (i.e., an event of MD).
As mentioned earlier, it is expected
that the performance degradation due to an event of MD
in the type-1 preamble detection
is worse than
that due to an event of FA.

Furthermore, we assume that the BS knows the number of
active type-2 devices, $K_2$.
Thus, we only consider the number of MD events,
which is the same\footnote{For example, suppose that $\cJ_2 = \{1,2,3\}$
when $\cL_2 = \{1,\ldots, 7\}$.  
If the index set of the detected preambles by the BS 
is $\{1,2,5\}$ (provided that $K_2 = 3$ is known),
the number of MD events is 1
(as the 3rd preamble is not detected) and the number of FA event
is also 1 (as the 5th preamble is incorrectly detected).}
as that of FA events, and present
the probability of MD to see the performance
of type-2 preamble detection.

\begin{figure}[thb]
\begin{center}
\includegraphics[width=\figwidth]{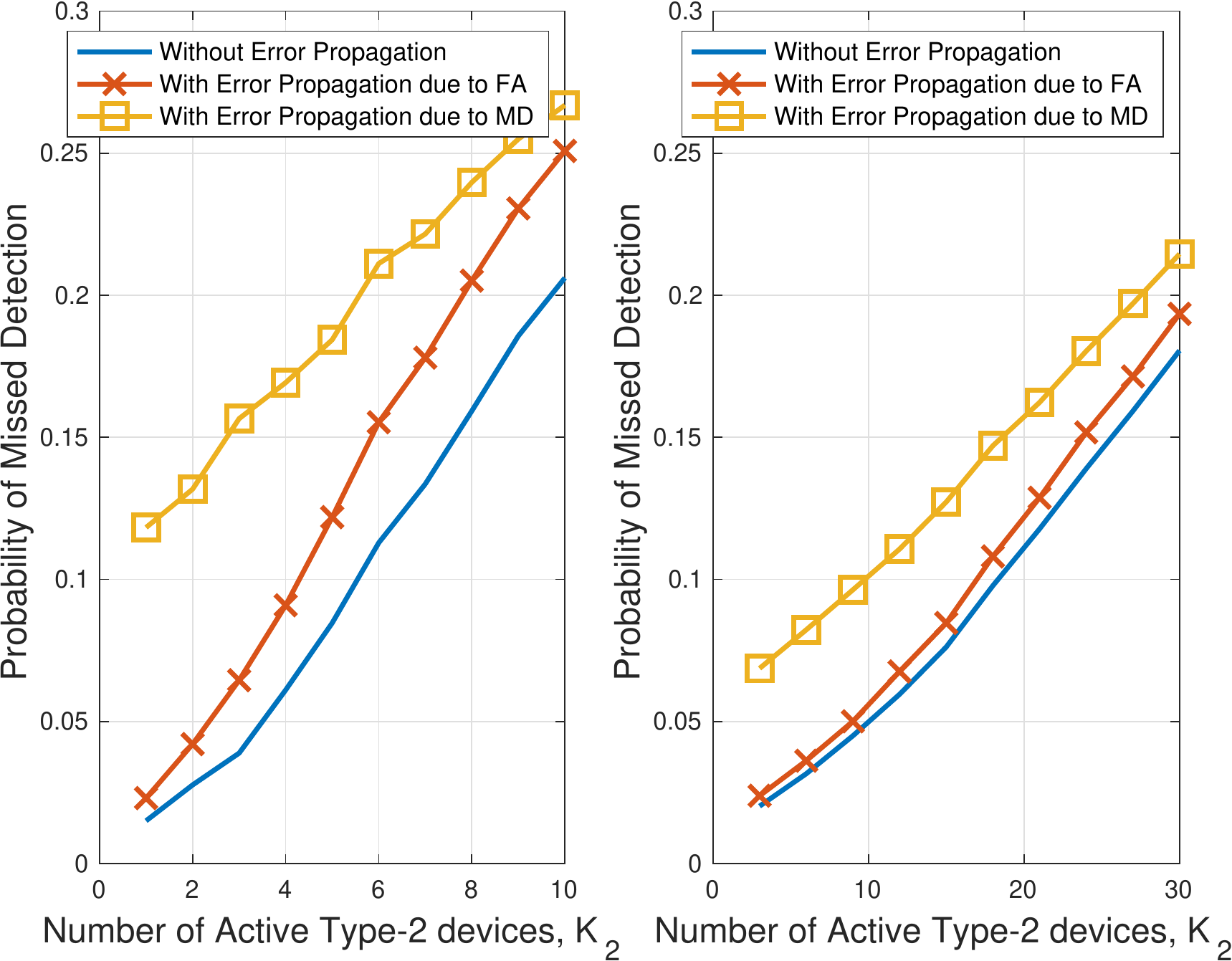}  \\
\hskip 0.5cm (a) \hskip 3.5cm (b)
\end{center}
\caption{The probabilities of MD of active type-2 
devices with/without error propagation
(due to FA and MD in the type-1 preamble detection)
as functions of the number of active type-2 devices, $K_2$,
when $M = 10$, $P_1 = 12$ dB, and $P_2 = 6$ dB:
(a) $N = L_1 = 13$ and $L_2 = 5N$; (b) $N = L_1 = 37$ and $L_2 = 5N$.}
        \label{Fig:t2_plt1}
\end{figure}

Fig.~\ref{Fig:t2_plt1}
shows the probabilities of MD of active type-2 
devices with/without error propagation
(due to FA and MD in the type-1 preamble detection)
as functions of the number of active type-2 devices, $K_2$,
when $M = 10$, $P_1 = 12$ dB, and $P_2 = 6$ dB.
In particular, in
Fig.~\ref{Fig:t2_plt1} (a),
the performance of
a small system with $N = L_1 = 13$ and
$L_2 = 5N$ is shown,
while in Fig.~\ref{Fig:t2_plt1} (b),
that of a large system with
$N = L_1 = 37$ and $L_2 = 5N$ is shown.
As expected, the performance degradation
due to an event of MD in 
the type-1 preamble detection 
is worse than that due to an event of FA.
We also see that a large system 
(i.e., $N = 37$) provides a better
performance than a small system (i.e., $N = 13$), which is 
well-known in the context of CS \cite{Eldar12}.
Note that this is not the case
of the type-1 preamble detection,
which was shown in Fig.~\ref{Fig:t1_plt2}.

\begin{figure}[thb]
\begin{center}
\includegraphics[width=\figwidth]{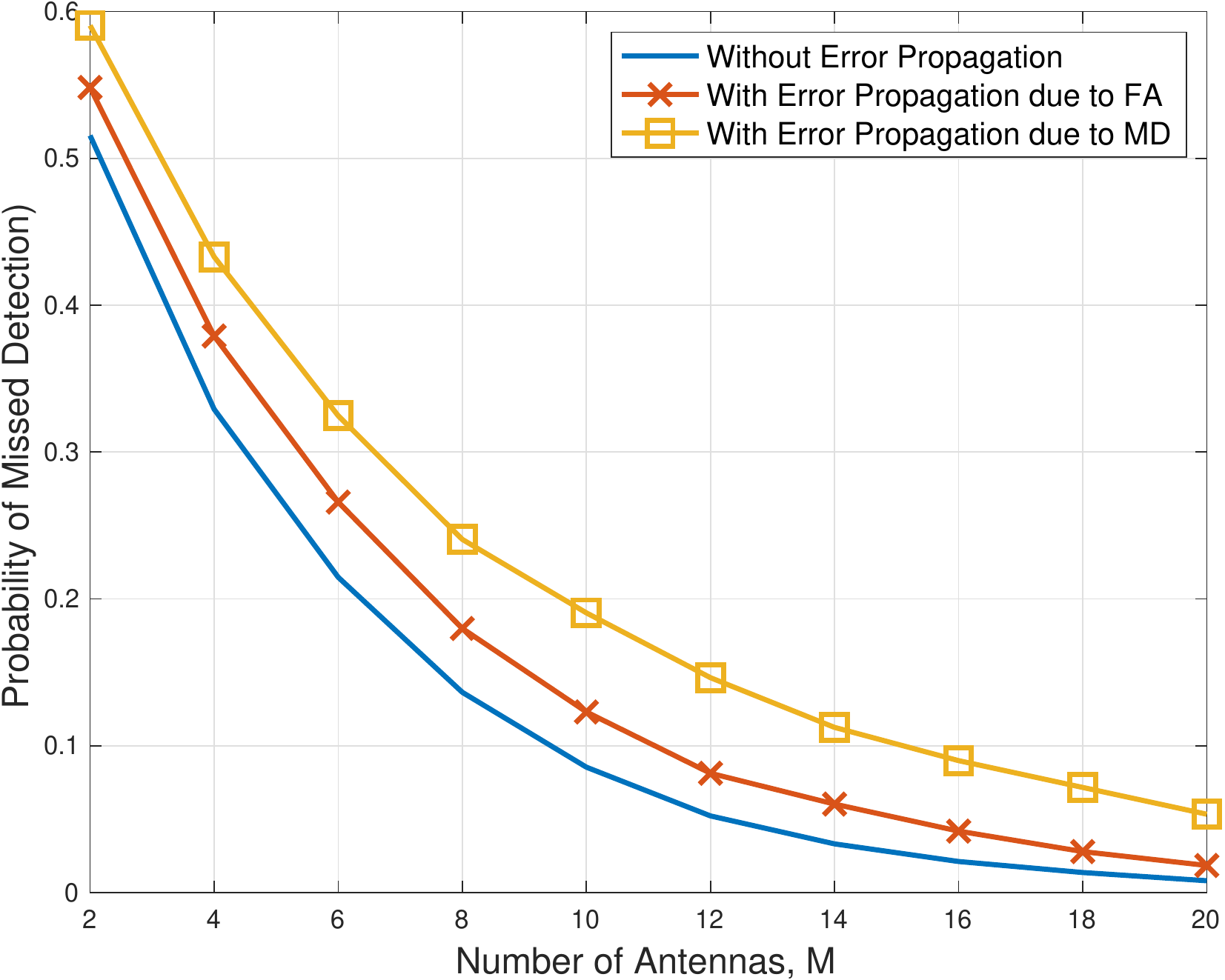}
\end{center}
\caption{The probabilities of MD of active type-2 
devices with/without error propagation
(due to FA and MD in the type-1 preamble detection)
as functions of the antennas at the BS, $M$, when $K_2 = 5$,
$N = L_1 = 13$, $L_2 = 5N$, $P_1 = 12$ dB, and $P_2 = 6$ dB.}
        \label{Fig:t2_plt2}
\end{figure}

In Fig.~\ref{Fig:t2_plt2},
the probabilities of MD of active type-2 
devices with/without error propagation
are shown 
as functions of the antennas at the BS, $M$, when $K_2 = 5$,
$N = L_1 = 13$, $L_2 = 5N$, $P_1 = 12$ dB, and $P_2 = 6$ dB.
Similar to Fig.~\ref{Fig:t1_plt2},
it is shown that a better performance can be achieved as
$M$ increases.

\begin{figure}[thb]
\begin{center}
\includegraphics[width=\figwidth]{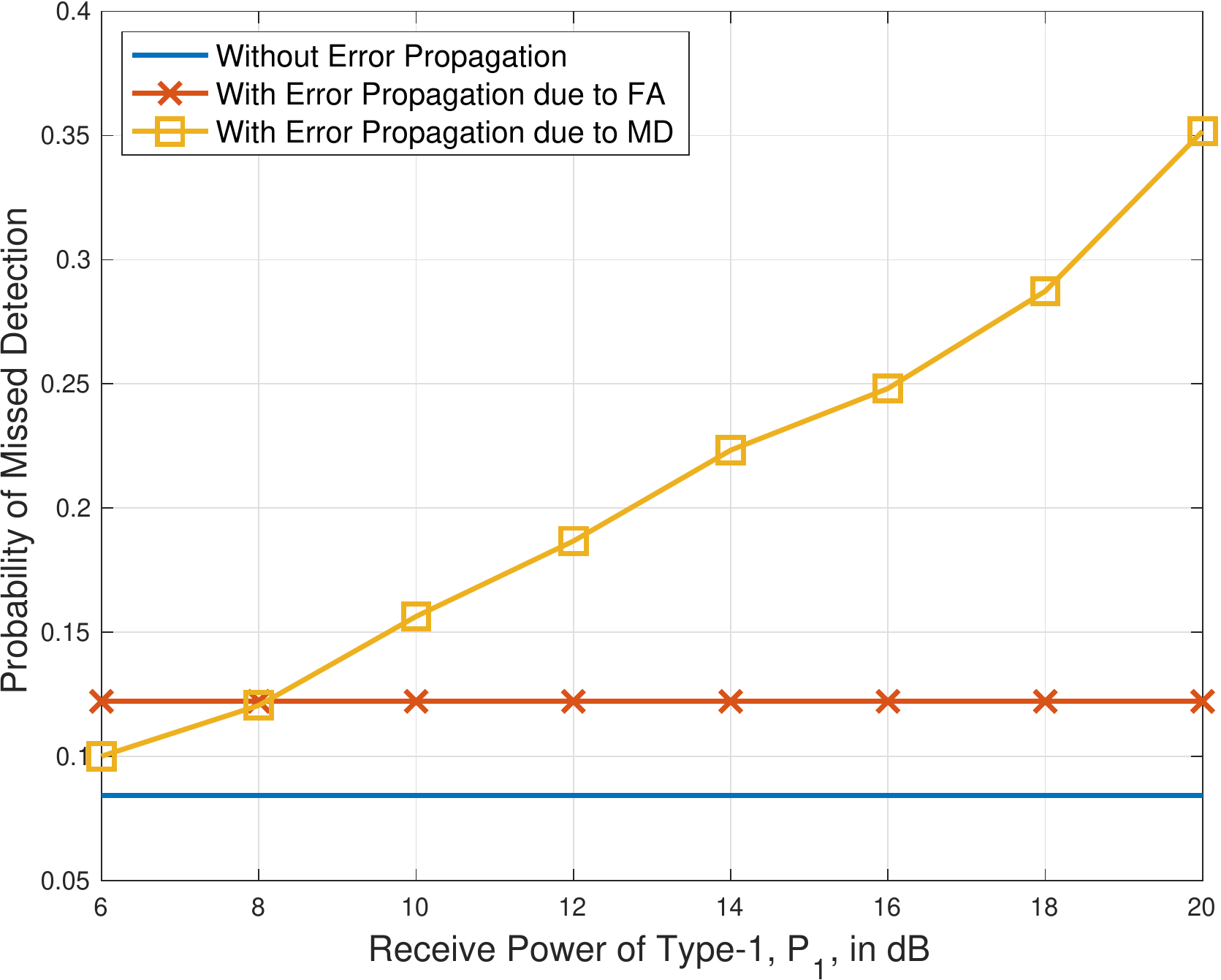}
\end{center}
\caption{The probabilities of MD of active type-2 
devices with/without error propagation
(due to FA and MD in the type-1 preamble detection)
as functions of the receive power
of type-1 devices, $P_1$, when $K_2 = 5$,
$N = L_1 = 13$, $L_2 = 5N$, $M = 10$, and $P_2 = 6$ dB.}
        \label{Fig:t2_plt3}
\end{figure}

Fig.~\ref{Fig:t2_plt3}
shows
the probabilities of MD of active type-2 
devices with/without error propagation
as functions of the receive power
of type-1 devices, $P_1$, when $K_2 = 5$,
$N = L_1 = 13$, $L_2 = 5N$, $M = 10$, and $P_2 = 6$ dB.
Clearly, the performance with error propagation 
due to an MD event in the type-1 preamble detection 
becomes worse as $P_1$ increases,
while that due to an FA event 
is independent of $P_1$ as any
signal in the subspace corresponding to
the related preamble in $\cL_1$ is suppressed.
This demonstrates that it is important to 
minimize the probability of MD
in the type-1 preamble detection
to keep a reasonable 
performance of type-2 preamble detection.

\begin{figure}[thb]
\begin{center}
\includegraphics[width=\figwidth]{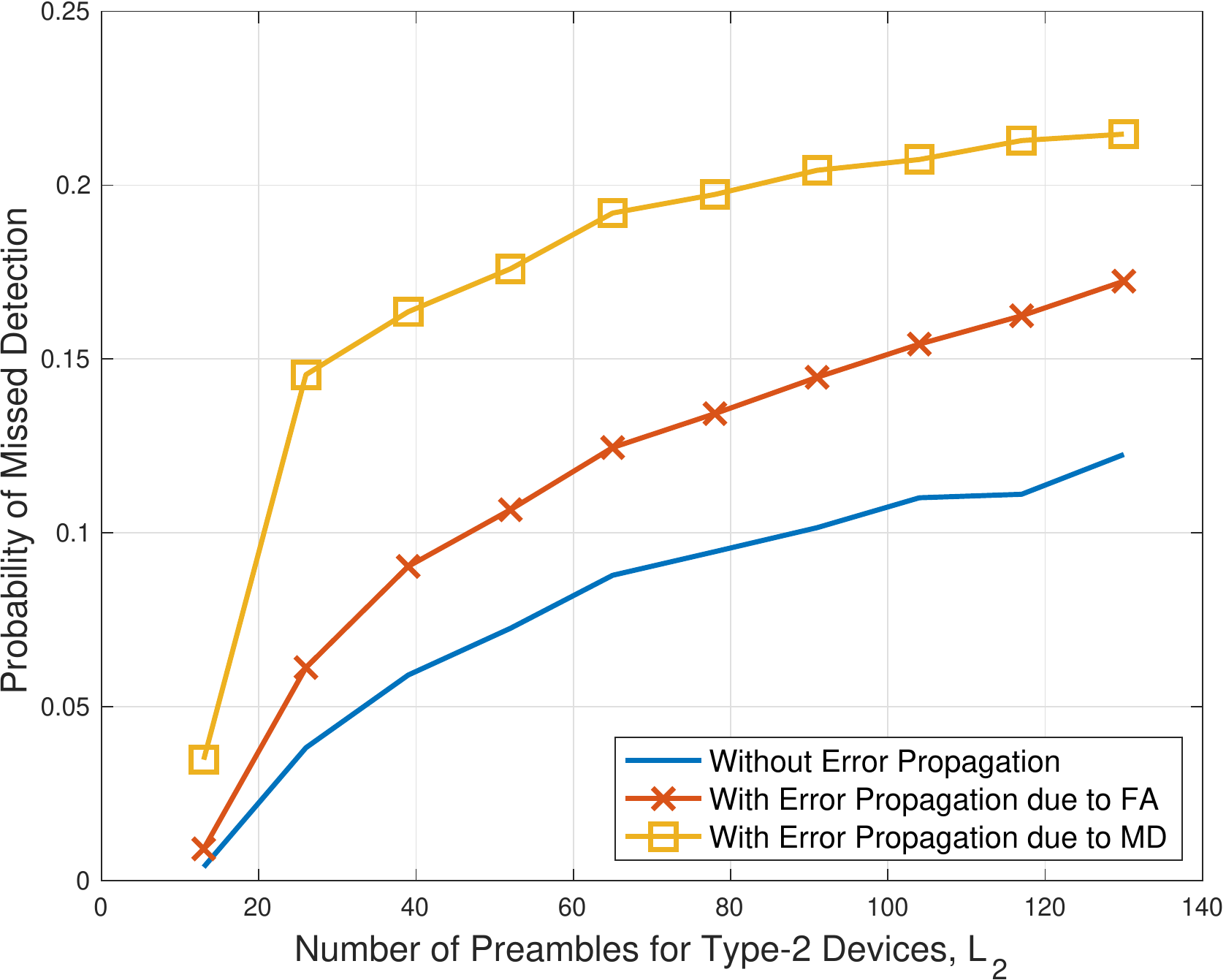}
\end{center}
\caption{The probabilities of MD of active type-2 
devices with/without error propagation
(due to FA and MD in the type-1 preamble detection)
as functions of the size of the preamble
pool for type-2 devices, $L_2 $, when $K_2 = 5$,
$N = L_1 = 13$, $M = 10$, $P_1 = 12$ dB, and $P_2 = 6$ dB.}
        \label{Fig:t2_plt4}
\end{figure}

It is expected that the probability of preamble collision
decreases with $L_2$ when $K_2$ is fixed.
However, if $L_2$ increases, the complexity
of the CAVI algorithm increases and its performance is also degraded.
To see the impact of $L_2$ on the performance
of the type-2 preamble detection,
we show the probabilities of MD of active type-2 
devices with/without error propagation
in Fig.~\ref{Fig:t2_plt4}
as functions of the size of the preamble
pool for type-2 devices, $L_2 $, when $K_2 = 5$,
$N = L_1 = 13$, $M = 10$, $P_1 = 12$ dB, and $P_2 = 6$ dB.
As expected, the probability of MD increases
with $L_2$. From this, we can see that
there is a trade-off between the probability of
preamble collision and the probability of MD, and 
$L_2$ should be chosen for
a balanced performance in terms of both probabilities.

From Figs.~\ref{Fig:t1_plt1} -~\ref{Fig:t2_plt4},
it can be shown that the probabilities of errors
(i.e., MD and FA) with orthogonal preambles
(for type-1 devices)
can be not only well predicted,
but also higher than those with non-orthogonal
preambles (for type-2 devices).
Thus, in RALP, $\cL_1$ can be used for delay-sensitive devices,
while $\cL_2$ for delay-tolerant devices.

\section{Concluding Remarks}	\label{S:Con}

In this paper, we proposed RALP using the notion
of power-domain NOMA to support two different types of devices,
namely type-1 devices (or 
delay-sensitive devices)
and  type-2 devices (or 
delay-tolerant devices)
with
one RB for high spectral efficiency.
Low-complexity detection methods have been studied
to detect transmitted preambles.
Thanks to the orthogonality of the preambles for
type-1 devices, it was possible to find
closed-form expressions for the probabilities of detection errors,
which can be used to determine
key parameters for target probabilities of errors.
This has been an important feature as a certain
performance guarantee 
can be ensured with known probabilities of errors for type-1
devices.

Since we mainly focused on RALP
in terms of the performance of the physical layer,
resource allocation and barring schemes with RALP
are not studied, which would be the topics
to be investigated in the future.

\bibliographystyle{ieeetr}
\bibliography{mtc}

\end{document}